\documentclass[namedreferences]{nato}
\usepackage{sprrefn}
\usepackage{txfonts}
\begin{document}
\begin{opening}
\title{Quantum computing with superconductors I: Architectures}
\author{Michael R. Geller\thanks{mgeller@uga.edu}}
\author{Emily J. Pritchett\thanks{epritchett@hal.physast.uga.edu}}
\institute{Department of Physics, University of Georgia, Athens, Georgia 30602, USA }
\author{Andrew T. Sornborger\thanks{ats@math.uga.edu}}
\institute{Department of Mathematics, University of Georgia, Athens, Georgia 30602, USA}
\author{F. K. Wilhelm\thanks{Present address: Physics Department and Insitute
for Quantum Computing, University of Waterloo, Waterloo, Ontario N2L 3G1, Canada; {\tt fwilhelm@iqc.ca}}}
\institute{Department Physik, Center for Nanoscience, and Arnold Sommerfeld Center for theoretical physics, Ludwig-Maximilians-Universit\"at, 80333 M\"unchen, Germany}
\runningtitle{Quantum computing with superconductors I}
\runningauthor{Geller {\it et al.}}

\begin{abstract}
Josephson junctions have demonstrated enormous potential as qubits for scalable quantum computing architectures. Here we discuss the current approaches for making multi-qubit circuits and performing quantum information processing with them.
\end{abstract}
\end{opening}

\tableofcontents

\section{Introduction}\label{introductionsection}

Macroscopic quantum behavior in a Josephson junction (JJ) was first demonstrated in the mid-1980's by John Clarke's group at UC Berkeley \cite{MartinisPRL85,DevoretPRL85,MartinisPRB87,ClarkeSci88}. These experiments used a superconducting device referred to as a large area, current-biased JJ, which would later become the phase qubit. Beginning in the mid-1990's the group of James Lukens at SUNY Stony Brook \cite{RousePRL95,FriedmanNat00} and a collaboration between the Delft University group of Hans Mooij and the MIT group of Terry Orlando \cite{MooijSci99,vanderWalSci00} demonstrated macroscopic quantum behavior in superconducting loops interrupted by one or more JJs (called superconducting quantum interference devices, or SQUIDS), what would later become flux qubits. And in the late-1990's the group of Yasunobu Nakamura at NEC in Tsukuba \cite{NakamuraPRL97,NakamuraNat99} developed the first Cooper-pair box or charge qubit. Many of the earlier experiments were motivated by seminal theoretical work of Caldeira and Leggett \cite{CaldeiraPRL81,Caldeira83}. 

The modern era of superconducting quantum computation began in 2002. That year, the group of Siyuan Han at the University of Kansas and the group of John Martinis, then at NIST Boulder and currently at UC Santa Barbara, independently showed that long-lived quantum states in a current-biassed JJ can be controllably prepared, manipulated, and subsequently measured \cite{YuSci02,MartinisPRL02}. This same year, the group of Michel Devoret, then at the CEA in Saclay and currently at Yale University, demonstrated similar quantum control using a Cooper-pair box \cite{VionSci02}. These experiments suggest that JJ-based qubits can be used as the building blocks of a solid-state quantum computer, creating a tremendous interest in this intrinsically scalable approach. An impressive list of additional experimental achievements soon followed, including the demonstration of two-qubit quantum logic \cite{YamamotoNat03}.

In this chapter we will review the current approaches for making multi-qubit systems. For a more detailed discussion of single qubits we refer to the excellent review by Makhlin, Sch\"on, and Shnirman \cite{MakhlinRMP01}. Also, a recent introductory account of the field has been given by You and Nori \cite{Youreview}. The approach we follow here is to construct circuit models for the basic qubits and coupled-qubit architectures. Many designs have been proposed, but only the simplest have been implemented experimentally to date. 

After reviewing in Sec.~\ref{basic qubits} the basic phase, flux, and charge qubits, we discuss three broad classes of coupling schemes. The simplest class uses fixed linear coupling elements, such as capacitors or inductors, and is discussed in Sec.~\ref{fixed coupling section}. The principal effect of fixed, weak couplings is to lift degeneracies of the uncoupled qubit pair. However, because such interactions are always present (always turned on), the uncoupled qubit states, which are often used as computational basis states, are not stationary. A variety of approaches have been proposed to overcome this shortcoming. In Sec.~\ref{tunable coupling section} we discuss tunable couplings that allow the interactions of Sec.~\ref{fixed coupling section} to be tuned, ideally between ``on" and ``off" values. A related class of {\it dynamic} couplings is discussed in Sec.~\ref{dynamic coupling section}, which make use of coupling elements that themselves have active internal degrees of freedom. They act like tunable coupling elements, but also have additional functionality coming from the ability to excite the internal degrees of freedom. Examples of this are resonator-based couplings, which we discuss in some detail.

\section{The basic qubits: phase, flux, and charge}\label{basic qubits}

\begin{figure}
\centerline{\includegraphics[width=5.0cm]{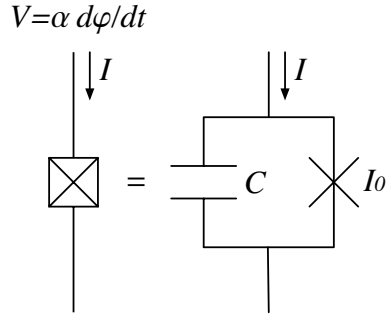}}
\caption{Circuit model for a current-biased JJ, neglecting dissipation. Here $\alpha \equiv \hbar/2e$.}
\label{jjcircuit figure}
\end{figure}

The primitive building block for all the qubits is the JJ shown in Fig.~\ref{jjcircuit figure}. The low-energy dynamics of this system is governed by the phase difference $\varphi$ between the condensate wave functions or order parameters on the two sides of the insulating barrier. The phase difference is an operator canonically conjugate to the Cooper-pair number difference $N$, according to\footnote{We define the momentum $P$ to be canonically conjugate to $\varphi$, and $N\equiv P/\hbar.$ In the phase representation, $N = -i \frac{\partial}{\partial \varphi}$.}
\begin{eqnarray}
[\varphi , N ] = i.
\label{canonical commutation relation}
\end{eqnarray}
The low-energy eigenstates $\psi_m(\varphi)$ of the JJ can be regarded as probability-amplitude distributions in $\varphi$. As will be explained below, the potential energy $U(\varphi)$ of the JJ is manipulated by applying a bias current $I$ to the junction, providing an external control of the quantum states $\psi_m(\varphi)$, including the qubit energy-level spacing $\Delta \epsilon$. The crossed box in Fig.~\ref{jjcircuit figure} represents a ``real" JJ. The cross alone represents a nonlinear element that satisfies the Josephson equations\footnote{$\alpha \equiv \hbar/2e$.}
\begin{eqnarray}
I = I_0 \sin \varphi \ \ \ \ \ \ \ \ {\rm and} \ \ \ \ \ \ \ \ V = \alpha {\dot \varphi},
\label{josephson equations}
\end{eqnarray}
with critical current $I_0$. The capacitor accounts for junction charging.\footnote{This provides a simple mean-field treatment of the {\it inter}-condensate electron-electron interaction neglected in the standard tunneling Hamiltonian formalism on which the Josephson equations are based.} A single JJ is characterized by two energy scales, the Josephson coupling energy
\begin{equation}
E_{\rm J} \equiv \frac{\hbar I_0}{2 e},
\end{equation}
where $e$ is the magnitude of the electron charge, and the Cooper-pair charging energy
\begin{equation}
E_{\rm c} \equiv {(2e)^2 \over 2C},
\end{equation}
with $C$ the junction capacitance. For example,
\begin{equation}
E_{\rm J} =  2.05 \, {\rm meV} \! \times \!  I_0[{\rm \mu A}]  
\ \ \ \ {\rm and} \ \ \ \ 
E_{\rm c} = {320 \, {\rm neV}  \over C[{\rm pF}]},
\end{equation}
where $I_0[{\rm \mu A}]$ and $C[{\rm pF}]$ are the critical current and junction capacitance in microamperes and picofarads, respectively. In the regimes of interest to quantum computation, $E_{\rm J}$ and $E_{\rm c}$ are assumed to be larger than the thermal energy $k_{\rm B}T$ but smaller than the superconducting energy gap $\Delta_{\rm sc},$ which is about $180 \, {\rm \mu eV}$ in Al. The relative size of $E_{\rm J}$ and $E_{\rm c}$ vary, depending on the specific qubit implementation.

\subsection{Phase qubit}

\begin{figure}
\centerline{\includegraphics[width=4.0cm]{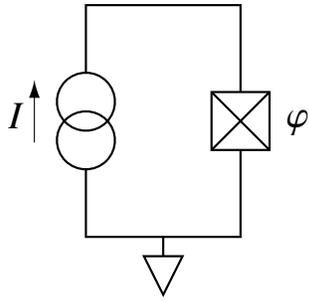}}
\caption{Basic phase qubit circuit.}
\label{phase qubit figure}
\end{figure}

\begin{figure}
\centerline{\includegraphics[width=9.0cm]{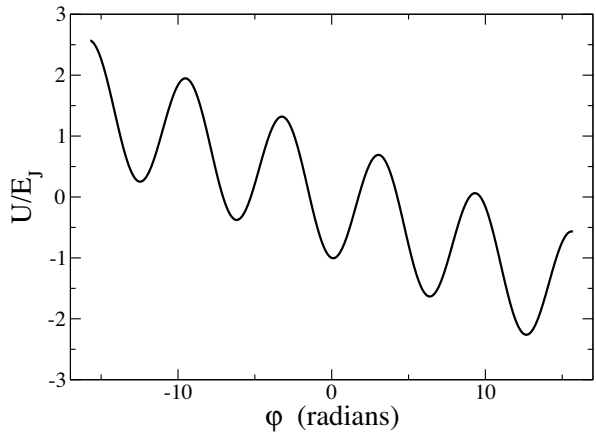}}
\caption{Effective potential for a current-biased JJ. The slope of the cosine potential is $s$. The potential is harmonic for the qubit states unless $s$ is very close to 1.}
\label{washboard figure}
\end{figure}

The basic phase qubit consists of a JJ with an external current bias, and is shown in Fig.~\ref{phase qubit figure}. The classical Lagrangian for this circuit is
\begin{eqnarray}
L_{\rm JJ} = {1 \over 2}M  {\dot \varphi}^2 - U, \ \ \ \ \ M \equiv {\hbar^2 \over 2 E_{\rm c}}.
\label{jj lagrangian}
\end{eqnarray}
Here
\begin{equation}
U \equiv - E_{\rm J} \big(\cos \varphi + s \, \varphi \big), \ \ \ \ {\rm with} \ \ \ \ s \equiv \frac{I}{I_0},
\label{effective potential}
\end{equation}
is the effective potential energy of the JJ, shown in Fig.~\ref{washboard figure}. Note that the ``mass" $M$ in (\ref{jj lagrangian}) actually has dimensions of ${\rm mass}  \times {\rm length}^2 \! .\, $ The form (\ref{jj lagrangian}) results from equating the sum of the currents flowing through the capacitor and ideal Josephson element to $I$. The phase qubit implementation uses $E_{\rm J} \gg E_{\rm c}.$

\begin{figure}
\centerline{\includegraphics[width=5.0cm]{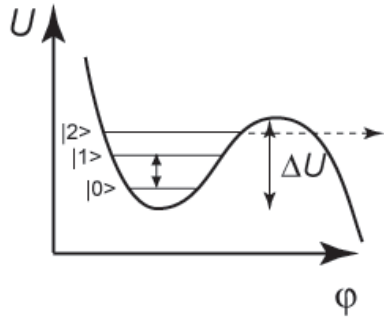}}
\caption{Effective potential in the anharmonic regime, with $s$ very close to 1. State preparation and readout are carried out in this regime.}
\label{cubic potential figure}
\end{figure}

According to the Josephson equations, the classical canonical momentum $P = \frac{\partial L}{\partial {\dot \varphi}}$ is proportional to the charge ${\sf Q}$ or to the number of Cooper pairs ${\sf Q}/2e$ on the capacitor according to $P = \hbar {\sf Q}/2e$. The quantum Hamiltonian can then be written as
\begin{equation}
H_{\rm JJ} = E_{\rm c} N^2 + U,
\label{jj hamiltonian}
\end{equation}
where $\varphi$ and $N$ are operators satisfying (\ref{canonical commutation relation}). Because $U$ depends on $s$, which itself depends on time, $H_{\rm JJ}$ is generally time-dependent. The low lying stationary states when $s \lesssim 1$ are shown in Fig.~\ref{cubic potential figure}. The two lowest eigenstates $|0\rangle$ and $|1\rangle$ are used to make a qubit. $\Delta \epsilon$ is the level spacing and $\Delta U$ is the height of the barrier.

A useful ``spin $\frac{1}{2}$" form of the phase qubit Hamiltonian follows by projecting (\ref{jj hamiltonian}) to the qubit subspace. There are two natural ways of doing this. The first is to use the basis of the $s$-dependent eigenstates, in which case
\begin{eqnarray}
H = - \frac{\hbar \omega_{\rm p}}{2} \, \sigma^z,
\end{eqnarray}
where
\begin{eqnarray}
\omega_{\rm p} \equiv \omega_{{\rm p}0} (1-s^2)^\frac{1}{4} \ \ \ \ {\rm and} \ \ \ \ \omega_{{\rm p}0} \equiv \sqrt{2 E_{\rm c} E_{\rm J}}/\hbar.
\end{eqnarray}
The $s$-dependent eigenstates are called instantaneous eigenstates, because $s$ is usually changing with time. The time-dependent Schr\"odinger equation in this basis contains additional terms coming from the time-dependence of the basis states themselves, which can be calculated in closed form in the harmonic limit \cite{Geller&ClelandPRA05}. These additional terms account for all nonadiabatic effects. 

The second spin form uses a basis of eigenstates with a fixed value of bias, $s_0$. In this case 
\begin{eqnarray}
H = - \frac{\hbar \omega_{\rm p}(s_0)}{2} \, \sigma^z - \frac{E_{\rm J} \ell}{\sqrt{2}} (s-s_0) \, \sigma^x,
\end{eqnarray}
where
\begin{eqnarray}
\ell \equiv \ell_0 (1-s_0)^{-\frac{1}{8}} \ \ \ \ {\rm and} \ \ \ \ \ell_0 \equiv \bigg(\frac{2 E_{\rm c}}{E_{\rm J}} \bigg)^\frac{1}{4}.
\label{l definition}
\end{eqnarray}
This form is restricted to $|s - s_0| \ll 1$, but it is very useful for describing rf pulses.

The angle $\ell$ characterizes the width of the eigenstates in $\varphi$. For example, in the $s_0$-eigenstate basis (and with $s_0$ in the harmonic regime), we have\footnote{$\sigma^0$ is the identity matrix.}
\begin{eqnarray}
\varphi = x_{01} \sigma^x + {\rm arcsin}(s_0) \, \sigma^0, \ \ \ \ {\rm with} \ \ \ \ x_{mm'} \equiv \langle m | \varphi | m' \rangle.
\end{eqnarray}
Here $x_{mm'}$ is an effective dipole moment (with dimensions of angle, not length), and $x_{01} = \ell/\sqrt{2}$.

\subsection{Charge qubit}

\begin{figure}
\centerline{\includegraphics[width=4.0cm]{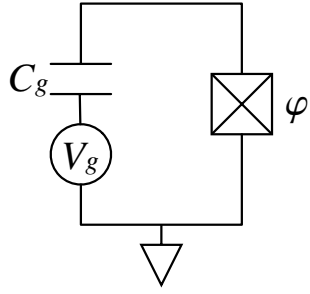}}
\caption{Basic charge qubit circuit. The upper wire constitutes the superconducting box or island.}
\label{charge qubit figure}
\end{figure}

In the charge qubit, the JJ current is provided capacitively, by changing the voltage $V_g$ on a gate, as in Fig.~\ref{charge qubit figure}. In this case $E_{\rm J} \ll E_{\rm c},$ and the small capacitance is achieved by using a Cooper-pair box, which is a nanoscale superconducting island or quantum dot.

The Lagrangian and Hamiltonian for this system are
\begin{eqnarray}
L = \frac{1}{2} \alpha^2 (C+C_{\rm g}){\dot \varphi}^2 + E_{\rm J} \cos \varphi - \alpha  C_{\rm g}V_{\rm g}  {\dot \varphi}
\end{eqnarray}
and
\begin{eqnarray}
H = E_{\rm c} (N-N_{\rm g})^2 - E_{\rm J} \cos \varphi, \ \ \ \ {\rm with} \ \ \ \ E_{\rm c} = \frac{(2e)^2}{2(C \! + \! C_{\rm g})}.
\label{charge qubit hamiltonian}
\end{eqnarray}
Here
\begin{eqnarray}
N_{\rm g} \equiv - \frac{C_{\rm g}V_{\rm g}}{2e}
\end{eqnarray}
is the gate charge, the charge qubit's control variable.

It is most convenient to use the charge representation here, defined by the Cooper-pair number eigenstates $|n\rangle$ satisfying
\begin{eqnarray}
N |n\rangle = n |n\rangle.
\end{eqnarray}
Because $e^{i \varphi}  |n\rangle =  |n+1\rangle$, the $\cos{\varphi}$ term in (\ref{charge qubit hamiltonian}) acts as a Cooper-pair tunneling operator. In the qubit subspace,
\begin{eqnarray}
N - N_{\rm g} &=& -(N_{\rm g}-{\textstyle{\frac{1}{2}}}) \sigma^0 - {\textstyle{\frac{1}{2}}} \sigma^z, \\
(N - N_{\rm g})^2 &=& (N_{\rm g}-{\textstyle{\frac{1}{2}}}) \sigma^z + {\rm const}, \\
\cos \varphi &=& {\textstyle{\frac{1}{2}}} \sigma^z.
\end{eqnarray}
The charge qubit Hamiltonian can then be written in spin form in the $\lbrace |0\rangle, |1\rangle \rbrace$ charge basis as
\begin{eqnarray}
H = E_{\rm c} (N_{\rm g} - {\textstyle{\frac{1}{2}}}) \sigma^z - \frac{E_{\rm J}}{2} \sigma^x,
\end{eqnarray}
or in the $\lbrace |+\rangle, |-\rangle \rbrace$ basis of $N_{\rm g}=\frac{1}{2}$ eigenstates
\begin{eqnarray}
| \pm\rangle \equiv \frac{|0\rangle \pm |1\rangle}{\sqrt{2}}
\end{eqnarray}
as
\begin{eqnarray}
H = E_{\rm c} (N_{\rm g} - {\textstyle{\frac{1}{2}}})\sigma^x - \frac{E_{\rm J}}{2} \sigma^z.
\end{eqnarray}

\subsection{Flux qubit}

\begin{figure}
\centerline{\includegraphics[width=5.0cm]{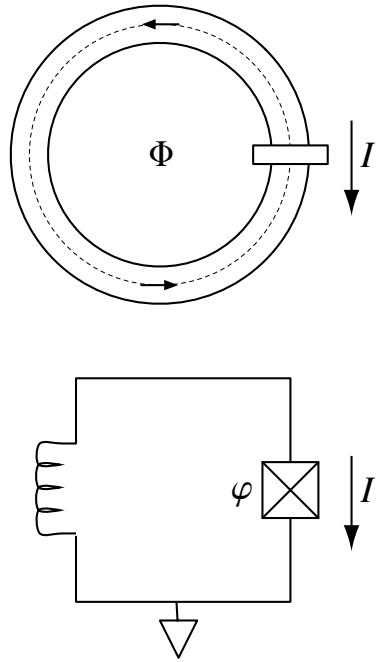}}
\caption{Basic rf-SQUID flux qubit and circuit model. $\Phi$ is the total flux threading the ring. The dashed curve in the upper figure indicates the integration contour $\Gamma$ used to derive condition (\ref{meissner condition}). The coil in the lower figure has self-inductance $L$.}
\label{flux qubit figure}
\end{figure} 

The flux qubit uses states of quantized circulation, or magnetic flux, in a SQUID ring. The geometry is illustrated in Fig.~\ref{flux qubit figure}. The current bias in this case is supplied by the circulating supercurrent. The total magnetic flux $\Phi$ can be written as
\begin{eqnarray}
\Phi = \Phi_{\rm x} - c LI,
\label{total flux condition}
\end{eqnarray}
where $\Phi_{\rm x}$ is the external contribution and $cLI$ is the self-induced component, with
\begin{eqnarray}
I= \alpha C {\ddot \varphi} +  I_0 \sin \varphi
\label{total current condition}
\end{eqnarray}
the circulating current and $L$ the self-inductance.\footnote{$L$ here is not to be confused with the Lagrangian.} The relations (\ref{total flux condition}) and (\ref{total current condition}) determine $\Phi$ given $\varphi$, but there is a second condition relating these quantities, namely
\begin{eqnarray}
\frac{\Phi}{\Phi_{\rm sc}} = \frac{\varphi}{2 \pi} \ {\rm mod} \, 1, \ \ \ \ \ {\rm with} \ \ \ \ \ \Phi_{\rm sc} \equiv \frac{hc}{2e}.
\label{meissner condition}
\end{eqnarray}
This second condition follows from the Meissner effect, which says that the current density in the interior of the ring vanishes, requiring the total vector potential ${\bf A}$ to be proportional to the gradient of the phase of the local order parameter. It is obtained by integrating ${\bf A}$ around the contour $\Gamma$ in Fig.~\ref{flux qubit figure}.

\begin{figure}
\centerline{\includegraphics[width=9.0cm]{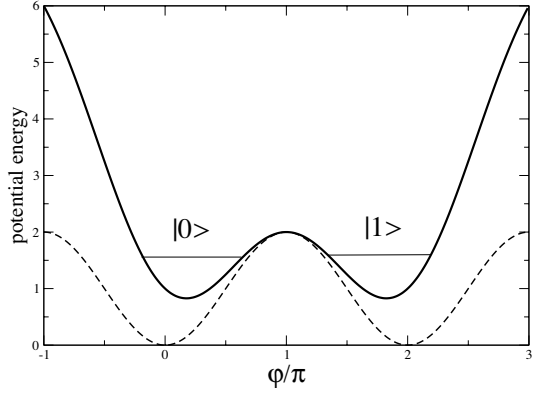}}
\caption{Double-well potential of the flux qubit. The dashed curve is the cosine potential of the JJ alone; the solid curve shows the modification caused by the self-inductance of the ring. The states $|0\rangle$ and $|1\rangle$ are that of circulating and counter-circulating supercurrent, which become degenerate  at the maximal frustration point (\ref{frustration condition}).}
\label{flux qubit potential figure}
\end{figure} 

The relation (\ref{total flux condition}) then becomes
\begin{eqnarray}
\alpha^2 C {\ddot \varphi} + E_{\rm J} \sin \varphi + \frac{\hbar^2 \omega_{\rm LC}^2}{2 E_{\rm c}}\bigg( \varphi - \frac{2\pi \Phi_{\rm x}}{\Phi_{\rm sc}} \bigg) = 0,
\end{eqnarray}
where
\begin{eqnarray}
\omega_{\rm LC} \equiv \frac{1}{\sqrt{LC}}.
\end{eqnarray}
This leads to the Lagrangian and Hamiltonian
\begin{eqnarray}
L = \frac{1}{2} \alpha^2 C {\dot \varphi}^2 + E_{\rm J} \cos \varphi - \frac{\hbar^2 \omega_{\rm LC}^2}{4 E_{\rm c}}\bigg( \varphi - \frac{2\pi \Phi_{\rm x}}{\Phi_{\rm sc}} \bigg)^2
\end{eqnarray}
and
\begin{eqnarray}
H = E_{\rm c} N^2 - E_{\rm J} \cos \varphi + \frac{\hbar^2 \omega_{\rm LC}^2}{4 E_{\rm c}}\bigg( \varphi - \frac{2\pi \Phi_{\rm x}}{\Phi_{\rm sc}} \bigg)^2.
\end{eqnarray}
The ring's self-inductance has added a quadratic contribution to the potential energy, centered at $2\pi \Phi_{\rm x}/\Phi_{\rm sc}$.

The control variable in the flux qubit is $\Phi_{\rm x}$. By choosing 
\begin{eqnarray}
\frac{\Phi_{\rm x}}{\Phi_{\rm sc}} = \frac{1}{2} \ {\rm mod} \ 1,
\label{frustration condition}
\end{eqnarray}
one produces the double-well potential shown in Fig.~\ref{flux qubit potential figure}. The condition (\ref{frustration condition}) corresponds of the point of maximum frustration between the two directions of circulating supercurrent. By deviating slightly from the point (\ref{frustration condition}), the energies of the $|0\rangle$ and $|1\rangle$ change, without changing the barrier height that controls the tunneling between the wells.

We can write the flux qubit Hamiltonian in spin form as
\begin{eqnarray}
H = B_z \sigma^z + B_x \sigma^x,
\end{eqnarray}
where $B_z$ and $B_x$ are parameters that depend on the SQUID geometry and $\Phi_{\rm x}$. In the simplest rf SQUID flux qubit discussed here, $B_z$ characterizes the well asymmetry, and is tunable (via $\Phi_{\rm x}$), whereas $B_x$ depends on the barrier height and is fixed by the value of $E_{\rm J}$. However, below we will describe a modification that allows the barrier height to be tuned as well.

Hybrid charge-flux qubits have also been demonstrated, and have shown to be successful in reducing decoherence caused by interactions with the environment \cite{VionSci02}.
 
\section{Fixed linear couplings}\label{fixed coupling section}

By fixed linear couplings we refer to coupling produced by electrically linear elements such as capacitors or inductors that lead to interaction Hamiltonians with fixed coupling strengths. In the cases usually considered, the coupling strengths are also weak, much smaller than the qubit level spacing, and we will assume that here as well. We discuss two prominent examples, capacitively coupled phase and charge qubits. For discussions of the third prominent example, inductively coupled flux qubits, we refer the reader to the literature \cite{MooijSci99,OrlandoPRB99,MakhlinRMP01,MassenvandenBrinkPRM05}.

\subsubsection{Capacitively coupled phase qubits}

\begin{figure}
\centerline{\includegraphics[width=6.0cm]{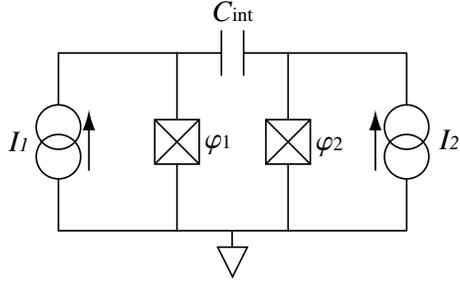}}
\caption{Capacitively coupled phase qubit circuit}
\label{coupled phase qubit figure}
\end{figure} 

Capacitively coupled phase qubits have been demonstrated by the University of Maryland group of Fred Wellstood \cite{BerkleySci03} and by the UC Santa Barbara group of John Martinis \cite{McDermottSci05}. The architecture was discussed theoretically by Johnson {\it et al.} \cite{JohnsonPRB03}, Blais {\it et al.} \cite{BlaisPRL03}, and Strauch {\it et al.} \cite{StrauchPRL03}.

Referring to Fig.~\ref{coupled phase qubit figure}, the equations of motion for the two phase variables are\footnote{$\alpha\equiv \hbar/2e$}
\begin{eqnarray}
\alpha^2 (C_1 + C_{\rm int}) {\ddot \varphi}_1 &+& E_{{\rm J}1}(\sin\varphi_1 - s_1) - \alpha^2 C_{\rm int} {\ddot \varphi}_2 = 0 \\
\alpha^2 (C_2 + C_{\rm int}) {\ddot \varphi}_2 &+& E_{{\rm J}2}(\sin\varphi_1 - s_1) - \alpha^2 C_{\rm int} {\ddot \varphi}_1 = 0, 
\end{eqnarray}
and the Lagrangian is
\begin{eqnarray}
L = \sum_i \big[{\textstyle{\frac{\alpha^2}{2}}} (C_i + C_{\rm int}) {\dot \varphi}_i^2 + E_{{\rm J}i} (\cos\varphi_i + s_i \varphi_i) \big] - \alpha^2 C_{\rm int} {\dot \varphi}_1 {\dot \varphi}_2.
\end{eqnarray}
To find the Hamiltonian, invert the capacitance matrix in
\begin{eqnarray}
\left( \begin{array}{c} p_1 \\p_2 \end{array} \right) = \alpha^2 
\left( \begin{array}{cc} C_1 + C_{\rm int} & -C_{\rm int} \\ -C_{\rm int} &   C_2 + C_{\rm int} \end{array} \right)\left( \begin{array}{c} {\dot \varphi}_1 \\ {\dot \varphi}_2 \end{array} \right), 
\end{eqnarray}
where the $p_i$ are the canonical momenta. This leads to
\begin{eqnarray}
H = \sum_i \bigg[ \frac{p_i^2}{2 \alpha^2 {\tilde C}_i} - E_{{\rm J}i} (\cos \varphi_i + s_i \varphi_i) \bigg] + \frac{p_1 p_2}{\alpha^2 {\tilde C}_{\rm int}},
\end{eqnarray}
where
\begin{eqnarray}
{\tilde C}_1 &\equiv& C_1 + \big( C_{\rm int}^{-1} + C_2^{-1} \big)^{-1}, \\
{\tilde C}_2 &\equiv& C_2 + \big( C_{\rm int}^{-1} + C_1^{-1} \big)^{-1}, \\
{\tilde C}_{\rm int} &\equiv& C_1 C_2 \big( C_1^{-1} + C_2^{-1} + C_{\rm int}^{-1} \big)^{-1} .
\end{eqnarray}
This can be written as
\begin{eqnarray}
H = \sum_i H_i + \delta H, \ \ \ \ \delta H \equiv g' N_1 N_2,
\end{eqnarray}
where
\begin{eqnarray}
g' \equiv \frac{(2e)^2}{{\tilde C}_{\rm int} } \rightarrow 2 \bigg(\frac{C_{\rm int}}{C}\bigg) E_{\rm c}.
\label{g' definition}
\end{eqnarray}
The arrow in (\ref{g' definition}) applies to the further simplified case of identical qubits and weak coupling.

The coupling constant $g'$ defined in in (\ref{g' definition}) is inconvenient, however, because the energy scale $E_{\rm c}$ appearing in (\ref{g' definition}) is too small. A better definition is
\begin{eqnarray}
g \equiv \frac{g'}{2 \ell_1 \ell_2} \rightarrow \bigg(\frac{C_{\rm int}}{C}\bigg) 
\hbar \omega_{\rm p} ,
\label{capacitively coupled phase qubit coupling constant definition} 
\end{eqnarray}
where $\ell$ is the scale introduced in (\ref{l definition}).

In the instantaneous basis, the spin form of the momentum operator is
\begin{eqnarray}
N = p_{01} \left( \begin{array}{cc} 0 & 1 \\ -1 & 0 \end{array} \right) , 
\end{eqnarray}
where
\begin{eqnarray}
p_{01} \equiv \langle 0 | p | 1 \rangle = - \frac{i}{\sqrt{2}\ell}.
\end{eqnarray}
Then
\begin{eqnarray}
H = \sum_i H_i + \delta H, \ \ \ \ \ H_i = - \frac{\hbar \omega_{\rm p}}{2} \sigma_i^z, \ \ \ \ \  \delta H \equiv g \sigma_i^y \sigma_i^y.
\end{eqnarray}
In the uncoupled qubit basis $\lbrace |00\rangle, |01\rangle, |10\rangle, |11\rangle \rbrace$, the qubit-qubit interaction in terms of (\ref{capacitively coupled phase qubit coupling constant definition}) is simply
\begin{eqnarray}
\delta H =g \left( \begin{array}{cccc} 0 & 0 & 0 & -1 \\ 0 & 0 & 1&  0 \\ 0 & 1 & 0 & 0 \\ -1 & 0 & 0 & 0 \end{array} \right).
\end{eqnarray}

Two-qubit quantum logic has not yet been demonstrated with this architecture. Methods for performing a controlled-Z and a modified swap gate have been proposed by Strauch {\it et al.} \cite{StrauchPRL03}, and four controlled-NOT implementations have also been proposed recently \cite{GelleretalPre06}. 

\subsubsection{Capacitively coupled charge qubits}

\begin{figure}
\centerline{\includegraphics[width=6.0cm]{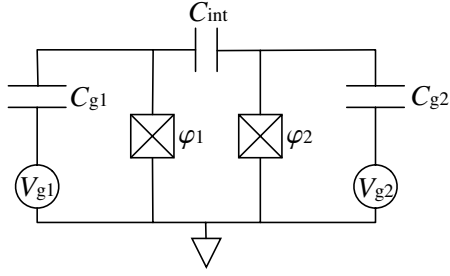}}
\caption{Capacitively coupled charge qubit circuit.}
\label{coupled quarge qubit figure}
\end{figure} 

A circuit for capacitively coupled charge qubits is given in Fig.~\ref{coupled quarge qubit figure}. This architecture has been demonstrated by Pashkin {\it et al.} \cite{PashkinNat03}, and used to perform a CNOT by Yamamoto {\it et al.} \cite{YamamotoNat03}. This work is currently the most advanced in the field of solid-state quantum information processing. The equations of motion for the two phases are\footnote{$\alpha\equiv \hbar/2e$}
\begin{eqnarray}
\alpha^2 (C_1 + C_{{\rm g}1} + C_{\rm int}) {\ddot \varphi}_1 &+& E_{{\rm J}1} \sin\varphi_1 - 
\alpha C_{{\rm g}1} {\dot V}_{{\rm g}1} - \alpha^2 C_{\rm int} {\ddot \varphi}_2 = 0 \\
\alpha^2 (C_2 + C_{{\rm g}1} + C_{\rm int}) {\ddot \varphi}_2 &+& E_{{\rm J}2} \sin\varphi_2 
-\alpha C_{{\rm g}2} {\dot V}_{{\rm g}2} - \alpha^2 C_{\rm int} {\ddot \varphi}_1 = 0, 
\end{eqnarray}
and the Lagrangian is
\begin{eqnarray}
L = \sum_i \big[{\textstyle{\frac{\alpha^2}{2}}} (C_i + C_{{\rm g}i} + C_{\rm int}) {\dot \varphi}_i^2 + E_{{\rm J}i} \cos\varphi_i  - \alpha C_{{\rm g}i}  V_{{\rm g}i} {\dot \varphi}_i \big] - \alpha^2 C_{\rm int} {\dot \varphi}_1 {\dot \varphi}_2.
\end{eqnarray}
Then the Hamiltonian is
\begin{eqnarray}
H &=& \sum_i \bigg[ \frac{(p_i + \alpha  C_{{\rm g}i}  V_{{\rm g}i})^2}{2 \alpha^2 {\tilde C}_i} - E_{{\rm J}i} \cos \varphi_i \bigg] \nonumber \\
&+& \frac{(p_1 + \alpha  C_{{\rm g}1} V_{{\rm g}1})(p_2 + \alpha  C_{{\rm g}2} V_{{\rm g}2})}{\alpha^2 {\tilde C}_{\rm int}},
\end{eqnarray}
where
\begin{eqnarray}
{\tilde C}_1 &\equiv& C_1 + C_{{\rm g}1} + \bigg[ C_{\rm int}^{-1} + \big(C_2 + C_{{\rm g}2} \big)^{-1} \bigg]^{-1}, \\
{\tilde C}_2 &\equiv& C_2 + C_{{\rm g}2} + \bigg[ C_{\rm int}^{-1} + \big(C_1 + C_{{\rm g}1} \big)^{-1} \bigg]^{-1}, \\
{\tilde C}_{\rm int} &\equiv& C_1 + C_{{\rm g}1} + C_2 + C_{{\rm g}2} + (C_1 + C_{{\rm g}1})(C_2 + C_{{\rm g}2}) C_{\rm int}^{-1}.
\end{eqnarray}
This can be written as
\begin{eqnarray}
H = \sum_i \big[ E_{{\rm c}i} (N_i - N_{{\rm g}i})^2 - E_{{\rm J}i} \cos \varphi_i \big] + \delta H,
\end{eqnarray} 
where
\begin{eqnarray}
\delta H = g (N_1 - N_{{\rm g}1})(N_2 - N_{{\rm g}2}), \ \ \ \ E_{{\rm c}i} \equiv \frac{(e2)^2}{2 {\tilde C}_i}, \ \ \ \ g \equiv \frac{(e2)^2}{2 {\tilde C}_{\rm int}}.
\end{eqnarray}

The spin form in the charge basis is
\begin{eqnarray}
H = \sum_i \big[ E_{{\rm c}i} (N_{{\rm g}i}-{\textstyle{\frac{1}{2}}})\sigma_i^z - \frac{E_{{\rm J}i}}{2} \sigma_i^x \big] + \delta H,
\end{eqnarray} 
with
\begin{eqnarray}
\delta H = \frac{g}{2}\big[ (N_{{\rm g}1}-{\textstyle{\frac{1}{2}}})\sigma_2^z +
(N_{{\rm g}2}-{\textstyle{\frac{1}{2}}})\sigma_1^z \big] + \frac{g}{4} \sigma_1^z \sigma_2^z.
\label{capacitively coupled charge qubit interaction}
\end{eqnarray}
When $N_{{\rm g}1} = N_{{\rm g}2} = {\textstyle{\frac{1}{2}}},$ this is a pure Ising interaction.

\section{Tunable couplings}\label{tunable coupling section}

By introducing more complicated coupling elements, we can introduce some degree of tunability into the architectures discussed above. 

\subsubsection{Tunable $E_{\rm J}$}

\begin{figure}
\centerline{\includegraphics[width=8.0cm]{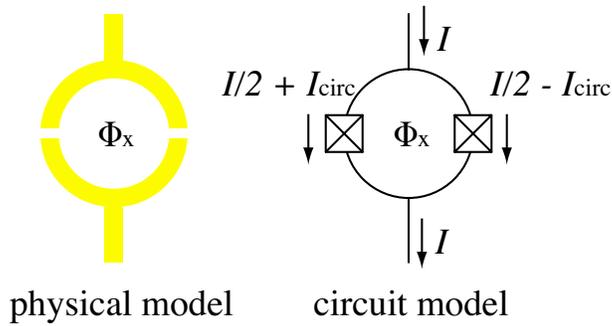}}
\caption{Tuning $E_{\rm J}$ with a dc SQUID.}
\label{tunable Ej figure}
\end{figure} 

A simple way to make the Josephson energy $E_{\rm J}$ effectively tunable in a circuit is to use a well known quantum interference effect in that occurs in a dc SQUID; see Fig.~\ref{tunable Ej figure}. The tunability of $E_{\rm J}$ can be understood from two different viewpoints.

The first is to imagine introducing a hole in a current-biased JJ as in the ``physical" model of Fig.~\ref{tunable Ej figure}. Tunneling occurs in the up and down direction in each of the left and right arms of the interferometer. Recalling our interpretation of $\cos \varphi$ as a Cooper-pair tunneling operator, the two arms of the interferometer result in 

\begin{eqnarray}
\cos \varphi \rightarrow \frac{e^{i(\varphi + \pi \frac{\Phi_{\rm x}}{\Phi_{\rm sc}})} 
+e^{-i(\varphi + \pi \frac{\Phi_{\rm x}}{\Phi_{\rm sc}})}}{2}
+ \frac{e^{i(\varphi - \pi \frac{\Phi_{\rm x}}{\Phi_{\rm sc}})} 
+e^{-i(\varphi - \pi \frac{\Phi_{\rm x}}{\Phi_{\rm sc}})}}{2}.
\end{eqnarray}
Here we have assumed a symmetric interferometer. The first pair of terms corresponds to tunneling (in both the up and down directions) in the left arm, which acquires half of the total Aharonov-Bohm phase $2 \pi \Phi_{\rm x} / \Phi_{\rm sc}$; the right arm has the opposite Aharonov-Bohm phase shift. Then the $\cos \varphi$ term in the potential energy of (\ref{jj hamiltonian}) becomes

\begin{eqnarray}
E_{\rm J}^0 \cos \varphi \rightarrow E_{\rm J}(\Phi_{\rm x}) \cos \varphi, \ \ \ \ {\rm with} \ \ \ \ 
E_{\rm J}(\Phi_{\rm x})  \equiv E_{\rm J}^0 \cos \bigg( \frac{\pi \Phi_{\rm x}}{\Phi_{\rm sc}} \bigg).
\label{tunable Ej}
\end{eqnarray}
The effective Josephson energy in (\ref{tunable Ej}) can be tuned by varying $\Phi_{\rm x}$.

The second way to obtain (\ref{tunable Ej}) is to consider the circuit model in Fig.~\ref{tunable Ej figure}, and again assume symmetry (identical JJs). This leads to the coupled equations of motion
\begin{eqnarray}
\alpha C {\ddot \varphi}_1 &+& I_0 \sin \varphi_1 = {\textstyle{\frac{I}{2}}} + I_{\rm circ}, \\
\alpha C {\ddot \varphi}_2 &+& I_0 \sin \varphi_2 = {\textstyle{\frac{I}{2}}} - I_{\rm circ} .
\end{eqnarray}

Defining 
\begin{eqnarray}
{\bar \varphi} \equiv \frac{\varphi_1 + \varphi_2}{2}
\end{eqnarray}
and using 
\begin{eqnarray}
\varphi_1 - \varphi_2 = \frac{2\pi \Phi_{\rm x}}{\Phi_{\rm sc}} 
\end{eqnarray}
then leads to
\begin{eqnarray}
\alpha^2 (2C) {\ddot {\bar \varphi}} + E_{\rm J}(\Phi_{\rm x}) \sin {\bar \varphi} - \alpha I = 0,
\end{eqnarray}
in agreement with (\ref{tunable Ej}). 

The ability to tune $E_{\rm J}$ is especially useful for inductively coupled flux qubits \cite{MakhlinRMP01}.

\subsubsection{Charge qubit register of Makhlin, Sch\"on, and Shnirman}

\begin{figure}
\centerline{\includegraphics[width=6.0cm]{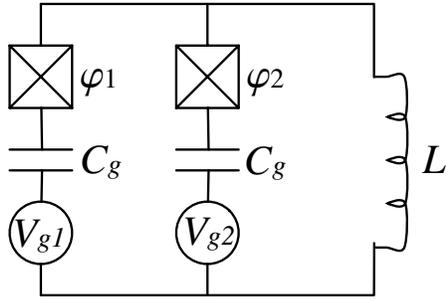}}
\caption{Circuit of Makhlin, Sch\"on, and Shnirman.}
\label{makhlin register figure}
\end{figure} 

Makhlin, Sch\"on, and Shnirman have proposed coupling charge qubits by placing them in parallel with an inductor, such that the resulting $LC$ oscillator (the capacitance provided by the JJs) has a frequency much higher than the qubit frequency \cite{MakhlinNat99}. The case of two qubits is illustrated in Fig.~\ref{makhlin register figure}, but the method applies to more than two qubits as well.

The derivation of the circuit Hamiltonian follows methods similar to that used above, and is
\begin{eqnarray}
H =  \sum_i \bigg[ E_{{\rm c}i} (N_{{\rm g}i} - {\textstyle{\frac{1}{2}}})\sigma_i^z - \frac{E_{{\rm J}i}}{2}\sigma_i^x \bigg] +  \frac{L C_{\rm qb}^2 E_{\rm J1} E_{\rm J2}}{4 \alpha^2 C^2}  \sigma_1^y \sigma_2^y .
\label{makhlin register hamiltonian}
\end{eqnarray}
The significant feature of the interaction in (\ref{makhlin register hamiltonian}), compared to (\ref{capacitively coupled charge qubit interaction}), is that the $E_{\rm J}$'s here can be tuned by using dc SQUIDs. This gives, in principle, a fully tunable interaction between any pair of qubits attached to the same inductor.

\subsubsection{Electrostatic transformer of Averin and Bruder}

\begin{figure}
\centerline{\includegraphics[width=8.0cm]{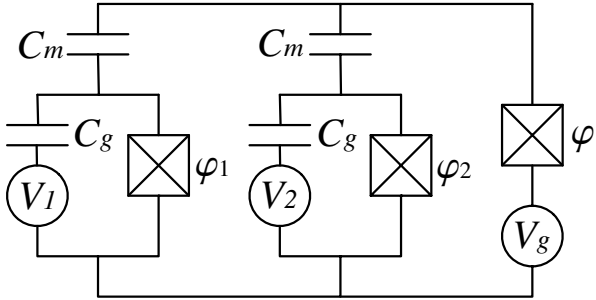}}
\caption{Electrostatic transformer.}
\label{transformer figure}
\end{figure} 

Averin and Bruder \cite{AverinPRL03} considered a related coupled charge qubit circuit, shown in Fig.~\ref{transformer figure}, which we have reorganized to emphasize the similarity to Fig.~\ref{makhlin register figure}. The Hamiltonian in this case is
\begin{eqnarray}
H = \sum_i \bigg[ E_{{\rm c}i} (N_i - N_{{\rm g}i})^2 - E_{{\rm J}i} \cos \varphi_i \bigg] + H_{\rm trans}
\end{eqnarray}
and
\begin{eqnarray}
H_{\rm trans} = E_{\rm c} (N - N_{\rm g} - q)^2 - E_{\rm J} \cos \varphi,
\end{eqnarray}
where
\begin{eqnarray}
q &\equiv&  q_{\rm g} - (N_1 - N_{\rm g1} + N_2 - N_{\rm g2})\frac{C_m}{C_{\Sigma i}} \\
q_{\rm g} &\equiv& 2 N_{\rm g} \bigg( 1 - \frac{C_m}{C_{\Sigma i}} \bigg).
\end{eqnarray}
The operator $q$ here is a function of the charge qubit variables, but commutes with the transformer degrees of freedom.

As in the register of Makhlin, Sch\"on, and Shnirman, we assume the transformer degrees of freedom are fast compared with the qubit variables, so that the transformer remains in its instantaneous ground state manifold. Then
\begin{eqnarray}
H_{\rm trans} \rightarrow \epsilon_0(q).
\end{eqnarray}
This finally leads to an effective Hamiltonian 
\begin{eqnarray}
H =  \sum_i \bigg[ E_{{\rm c}i} (N_{{\rm g}i} - {\textstyle{\frac{1}{2}}})\sigma_i^z - \frac{E_{{\rm J}i}}{2}\sigma_i^x \bigg] +  \sum_i a \sigma_i^z + b \sigma_1^z \sigma_2^z ,
\end{eqnarray}
involving charge qubit variables only, where
\begin{eqnarray}
a &\equiv& \frac{\epsilon_0(q_0 +  \frac{C_m}{C_{\Sigma i}} ) - \epsilon_0(q_0 - \frac{C_m}{C_{\Sigma i}} )}{4} , \\
b &\equiv& \frac{\epsilon_0(q_0 +  \frac{C_m}{C_{\Sigma i}} ) + \epsilon_0(q_0 - \frac{C_m}{C_{\Sigma i}} ) - 2 \epsilon_0(q_0)}{4} .
\end{eqnarray}
The discrete second-order derivative $b$, which can be interpreted as a capacitance, can be tuned to zero by varying $q_0$, providing the desired tunability.

\subsection{RF coupling}

Finally, we briefly mention an interesting proposal by Rigetti, Blais, and Devoret, to use rf pulses to effectively bring permanently detuned qubits into resonance \cite{RigettiPRL05}. This is a very promising approach, but has not yet been demonstrated experimentally. 

\section{Dynamic couplings: Resonator coupled qubits}\label{dynamic coupling section}

Several investigators have proposed the use of $LC$ resonators \cite{ShnirmanPRL97,MakhlinNat99,MooijSci99,YouPRL02,Yukon02,BlaisPRL03,PlastinaPRB03,ZhouPRA04}, superconducting cavities \cite{BlaisPRA04,WallraffNat04}, or other types of oscillators \cite{MarquardtPRB01,ZhuPRA03} to couple JJs together. Although harmonic oscillators are ineffective as computational qubits, because the lowest pair of levels cannot be frequency selected by an external driving field, they are quite desirable as bus qubits or coupling elements. Resonators provide for additional functionality in the coupling, and can be made to have very high $Q$ factor. Here we will focus on phase qubits coupled by nanomechanical resonators \cite{Cleland&GellerPRL04,Geller&ClelandPRA05,SornborgeretalPRA04,Pritchett&GellerPRA05}.

\subsection{Qubit-resonator Hamiltonian}

\begin{figure}
\centerline{\includegraphics[width=10.0cm]{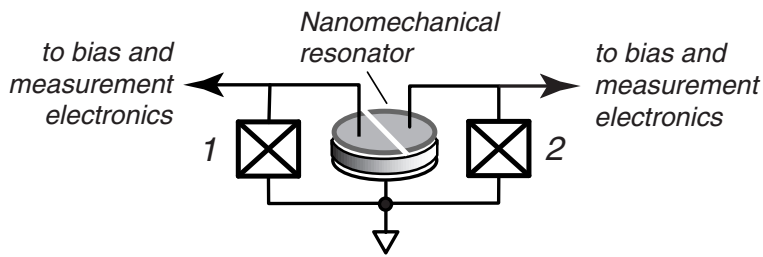}}
\caption{Two current-biased Josephson junctions (crossed boxes) coupled to a piezoelectric disc resonator.}
\label{nems coupling figure}
\end{figure} 

The Hamiltonian that describes the low-energy dynamics of a single large-area, current-biased JJ, coupled to a piezoelectric nanoelectromechanical disk resonator, can be written as \cite{Cleland&GellerPRL04,Geller&ClelandPRA05}
\begin{equation}
H = \sum_m \epsilon_m c_m^\dagger c_m + \hbar \omega_0 a^\dagger a - i g \sum_{mm'} x_{mm'} c_m^\dagger c_{m'} (a-a^\dagger),
\label{full Hamiltonian}
\end{equation}
where the $\{c_m^\dagger \}$ and $\{c_m\}$ denote particle creation and annihilation operators for the Josephson junction states $(m \! = \! 0,1,2,\dots), a$ and $a^\dagger$ denote ladder operators for the phonon states of the resonator's dilatational (thickness oscillation) mode of frequency $\omega_0$, $g$ is a coupling constant with dimensions of energy, and $x_{mm'} \equiv \langle m | \varphi | m' \rangle$. The value of $g$ depends on material properties and size of the resonator, and can be designed to achieve a wide range of values. An illustration showing two phase qubits coupled to the same resonator is given in Fig.~\ref{nems coupling figure}. Interactions between the JJ and resonator may be controlled by changing the JJ current, giving rise to changes in the JJ energy spacing, $\Delta\epsilon$. For instance, a state can be transferred from the JJ to the resonator by bringing the JJ and resonator in resonance, $\Delta\epsilon = \hbar\omega_0$, and waiting for a specified period.

\subsection{Strong coupling and the RWA}

For small couplings $g \ll \Delta\epsilon$, the JJ-resonator system may be approximated by the Jaynes-Cummings model; this is usually referred to as the rotating wave approximation (RWA). However, once the coupling becomes comparable to the level spacing, $g \approx \Delta\epsilon$, the RWA breaks down. When the JJ is weakly coupled to the resonator, with $g/\Delta\epsilon$ below a few percent, gates such as a memory operation (state transfer to and from the resonator) work well, and qubits are stored and retrieved with high fidelity. {\it However, such gates are intrinsically slow}. As $g/\Delta\epsilon$ is increased, making the gate faster, the fidelity becomes very poor, and it becomes necessary to deviate from the RWA protocol. Below, we first discuss an analytical approach to capture the leading corrections to the RWA at intermediate coupling strengths \cite{SornborgeretalPRA04}. We then discuss a strong coupling information processing example: a quantum memory register \cite{Pritchett&GellerPRA05}.

\subsection{Beyond the RWA}

For simplicity we will consider only two levels in a single junction. However, all possible phonon-number states are included. The Hamiltonian may then be written as the sum of two terms, $H = H_{\rm JC} + V$. The first term, 
\begin{eqnarray}
H_{\rm JC} \equiv  \epsilon_0 \, c_0^\dagger c_0+ \epsilon_1 \, c_1^\dagger c_1 + \hbar \omega_0 \, a^\dagger a 
- i g x_{01} [ c_1^\dagger c_0 a - c_0^\dagger c_1 a^\dagger],
\end{eqnarray}
is the exactly solvable Jaynes-Cummings Hamiltonian, the eigenfunctions of which are known as dressed states. We will consider the second term, 
\begin{eqnarray}
V \equiv  - i g \big[ x_{00}  c_0^\dagger c_0 (a-a^\dagger) +  x_{01} c_0^\dagger c_1 a  
- x_{01} c_1^\dagger c_0 a^\dagger + x_{11}  c_1^\dagger c_1 (a-a^\dagger)\big],
\end{eqnarray}
as a perturbation. The RWA applied to the Hamiltonian $H$ amounts to neglecting $V$. Therefore, perturbatively including $V$ is equivalent to perturbatively going beyond the RWA.

\subsubsection{Dressed states}

The eigenstates of $H_{\rm JC}$, or the dressed states, are labeled by the nonnegative integers $j=0,1,2,\dots$ and a sign $\sigma= \pm 1$. On resonance, these are
\begin{equation}
|\psi^\sigma_j\rangle \equiv {|0,j+1\rangle - i \sigma | 1,j\rangle \over \sqrt{2}}, \ \ \ \ \ (\omega_{\rm d}=0)
\label{resonant JC eigenstates}
\end{equation}
and
\begin{equation}
W^\sigma_j \equiv  \epsilon_0 + (j+1) \hbar \omega_0 + \sigma \sqrt{j+1} \ { \hbar \Omega_0(0) \over 2}. \ \ \ \ \ (\omega_{\rm d}=0)
\label{resonant JC energies}
\end{equation}
Here, the vacuum $(j \! = \! 0)$ Rabi frequency on resonance is $\Omega_0(0) = 2 g |x_{01}|/\hbar$.

\subsubsection{Dressed state propagator}

In quantum computing applications one will often be interested in calculating transition amplitudes of the form
\begin{equation}
\langle {\rm f} | e^{-i H t/\hbar} | {\rm i} \rangle,
\end{equation}
where $| {\rm i} \rangle$ and $| {\rm f} \rangle$ are arbitrary initial and final states of the uncoupled qubit-resonator system. Expanding $| {\rm i} \rangle$ and $| {\rm f} \rangle$ in the dressed-state basis reduces the time-evolution problem to that of calculating the quantity
\begin{equation}
G_{jj'}^{\sigma \sigma'} \! (t) \equiv \langle \psi_{j}^{\sigma} |e^{-i H t/\hbar} |\psi_{j'}^{\sigma'}\rangle,
\label{propagator definition}
\end{equation}
as well as  $\langle \psi_{j}^{\sigma} |e^{-i H t/\hbar} | 00 \rangle$ and $\langle 00 |e^{-i H t/\hbar} |00 \rangle$. 
$ \ G_{jj'}^{\sigma \sigma'} \! (t)$ is a propagator in the dressed-state basis, and would be equal to $\delta_{\sigma \sigma'} \delta_{jj'} e^{-i W_j^\sigma t/\hbar}$ if $V$ were absent, that is, in the RWA.

To be specific, we imagine preparing the system at $t=0$ in the state $|10\rangle$, which corresponds to the qubit in the excited state $m=1$ and the resonator in the ground state $n=0$. We then calculate the interaction-representation probability amplitude 
\begin{equation}
c_{mn}(t) \equiv e^{i E_{mn}t/\hbar} \langle mn| e^{-i H t/\hbar} |10\rangle
\end{equation}
for the system at a later time $t$ to be in the state $|mn\rangle$. Here $E_{mn} \equiv \epsilon_m + n \hbar \omega_0$. Inserting complete sets of the dressed states leads to
\begin{equation}
c_{00}(t) = \sum_{\sigma j} \langle \psi_j^\sigma |10\rangle \langle 00 | e^{-i H t/\hbar} |\psi_{j}^{\sigma}\rangle,
\end{equation}
and, for $mn \neq 00$,

\begin{equation}
c_{mn}(t) = e^{i E_{mn} t/\hbar} \sum_{j=0}^\infty 
\left( \begin{array}{c} \langle \psi_j^+ |mn \rangle \\  \langle \psi_j^- |mn \rangle \end{array}\right)^\dagger \left( \begin{array}{cc} G_{j0}^{++} & G_{j0}^{+-} \\ G_{j0}^{-+} & G_{j0}^{--} \end{array} \right) 
\left( \begin{array}{c} \langle \psi_0^+ |10 \rangle \\ \langle \psi_0^- |10 \rangle \end{array}\right).
\label{cmn equation}
\end{equation}
So far everything is exact within the model defined in Eq.~(\ref{full Hamiltonian}).

To proceed, we expand the dressed-state propagator in a basis of {\it exact} eigenstates $|\Psi_\alpha \rangle$ of $H$, leading to
\begin{equation}
G_{jj'}^{\sigma \sigma'} \! (t) = \sum_\alpha  \langle \psi_j^\sigma | \Psi_\alpha \rangle \, \langle \psi_{j'}^{\sigma'} | \Psi_\alpha \rangle^* \, e^{-i {\cal E}_\alpha t/\hbar}.
\label{propagator eigenfunction expansion}
\end{equation}
Here ${\cal E}_\alpha$ is the energy of stationary state $|\Psi_\alpha\rangle$. The propagator is an infinite sum of periodic functions of time. We approximate this quantity by evaluating the $|\Psi_\alpha\rangle$ and  ${\cal E}_\alpha$ perturbatively in the dressed-state basis.

We test our perturbed dressed-state method for the case of a finite-dimensional single-qubit, five-phonon system. The bias current is chosen to make the system exactly in resonance. The Hamiltonian for this system is diagonalized numerically, and the probability amplitudes $c_{mn}(t)$ are calculated exactly, providing a test of the accuracy of the analytic perturbative solutions. Setting the initial state to be $c_{mn}(0) = \delta_{m1} \delta_{n0}$, as assumed previously, we simulate the transfer of a qubit from the Josephson junction to the resonator, by leaving the systems in resonance for half a vacuum Rabi period $\pi \hbar / g |x_{01}|.$

\begin{figure}
\centerline{\includegraphics[width=9.0cm]{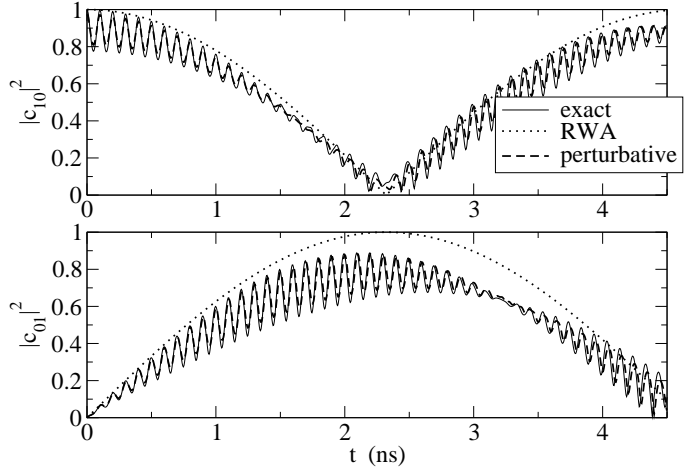}}
\caption{Probabilities $|c_{10}|^2$ and $|c_{01}|^2$ for the intermediate case of $g/\Delta \epsilon = 0.30$. Here there are large deviations from the RWA behavior, which are correctly accounted for by the dressed-state perturbative method.}
\label{g=0.30 figure}
\end{figure}

In Fig.~\ref{g=0.30 figure}, we plot the probabilities for a relatively strong coupling, $g/\Delta \epsilon = 0.30$. For this coupling strength, the RWA is observed to fail. For example, the RWA predicts a perfect state transfer between the junction and the resonator, and does not exhibit the oscillations present in the exact solution. The dressed state perturbation theory does correctly capture these oscillations.

\subsection{Memory operation with strong coupling}

Here we study a complete memory operation, where the qubit is stored in the resonator and then transferred back to the JJ, for a large range of JJ-resonator coupling strengths \cite{Pritchett&GellerPRA05}. Also, we show that a dramatic improvement in memory performance can be obtained by a numerical optimization procedure where the resonant interaction times and off-resonant detunings are varied to maximize the overall gate fidelity. This allows larger JJ-resonator couplings to be used, leading to faster gates and therefore more operations carried out within the available coherence time. The results suggest that it should be possible to demonstrate a fast quantum memory using existing superconducting circuits, which would be a significant accomplishment in solid-state quantum computation. 

\begin{figure}
\centerline{\includegraphics[width=8.0cm]{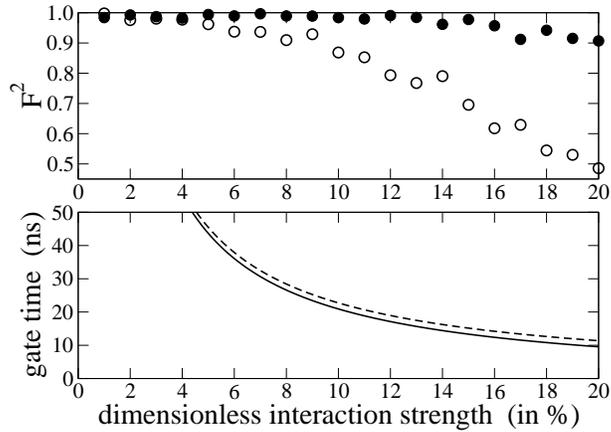}}
\caption{(upper panel) Memory fidelity for equator state $2^{-{1\over2}}(|0\rangle+|1\rangle)$ as a function of $g/\Delta\epsilon$, using both the RWA (unfilled circles) and optimized (solid circles) pulse times. (lower panel) Time needed to store and retrieve state, using both the RWA (dashed curve) and optimized (solid curve) pulse times.}
\label{Fvsg figure}
\end{figure}

In the upper panel of Fig.~\ref{Fvsg figure} we plot the memory fidelity for the qubit state $2^{-{1 \over 2}}(|0\rangle+|1\rangle)$ as a function of $g/\Delta\epsilon.$ We actually report the fidelity squared, 
\begin{equation}
F^2 = \big| \alpha^* c_{00}(t_{\rm f}) + \beta^* c_{10}(t_{\rm f}) \big|^2 \! ,
\label{F definition}
\end{equation}
which is the probability that the memory device operates correctly.  As expected, the fidelity gradually decreases with increasing $g.$  The lower panel of Fig.~\ref{Fvsg figure} gives the gate time as a function of $g/\Delta\epsilon.$ These results suggest that memory fidelities better than 90\% can be achieved using phase qubits and resonators with coherence times longer than a few tens of ns.

\begin{acknowledgements}

This work was supported by the NSF under grants DMR-0093217 and CMS-040403.

\end{acknowledgements}

\bibliography{/Users/mgeller/Papers/bibliographies/MRGpre,/Users/mgeller/Papers/bibliographies/MRGbooks,/Users/mgeller/Papers/bibliographies/MRGgroup,/Users/mgeller/Papers/bibliographies/MRGphonons,/Users/mgeller/Papers/bibliographies/MRGcm,/Users/mgeller/Papers/bibliographies/MRGnano,/Users/mgeller/Papers/bibliographies/MRGqc-josephson,/Users/mgeller/Papers/bibliographies/MRGqc-architectures,/Users/mgeller/Papers/bibliographies/MRGqc-general,/Users/mgeller/Papers/bibliographies/MRGqc-reviews,natonotes}

\begin{thebibliography}{}

\bibitem[\protect\citeauthoryear{Averin and Bruder}{2003}]{AverinPRL03}
Averin, D.~V. and C. Bruder: 2003, `Variable electrostatic transformer:
  {C}ontrollable coupling of two charge qubits'.
\newblock {\em Phys. Rev. Lett.} {\bf 91}, 57003.

\bibitem[\protect\citeauthoryear{Berkley et~al.}{2003}]{BerkleySci03}
Berkley, A.~J., H. Xu, R.~C. Ramos, M.~A. Gubrud, F.~W. Strauch, P.~R. Johnson,
  J.~R. Anderson, A.~J. Dragt, C.~J. Lobb, and F.~C. Wellstood: 2003,
  `Entangled macroscopic quantum states in two superconducting qubits'.
\newblock {\em Science} {\bf 300}, 1548--50.

\bibitem[\protect\citeauthoryear{Blais et~al.}{2004}]{BlaisPRA04}
Blais, A., R.-S. Huang, A. Wallraff, S.~M. Girvin, and R.~J. Schoelkopf: 2004,
  `Cavity quantum electrodynamics for superconducting electrical circuits: {A}n
  architecture for quantum computation'.
\newblock {\em Phys. Rev. A} {\bf 69}, 62320.

\bibitem[\protect\citeauthoryear{Blais et~al.}{2003}]{BlaisPRL03}
Blais, A., A. Massen van~den Brink, and A.~M. Zagoskin: 2003, `Tunable coupling
  of superconducting qubits'.
\newblock {\em Phys. Rev. Lett.} {\bf 90}, 127901.

\bibitem[\protect\citeauthoryear{Caldeira and Leggett}{1981}]{CaldeiraPRL81}
Caldeira, A.~O. and A.~J. Leggett: 1981, `Influence of dissipation on quantum
  tunneling in macroscopic systems'.
\newblock {\em Phys. Rev. Lett.} {\bf 46}, 211--4.

\bibitem[\protect\citeauthoryear{Caldeira and Leggett}{1983}]{Caldeira83}
Caldeira, A.~O. and A.~J. Leggett: 1983, `Quantum tunneling in a dissipative
  system'.
\newblock {\em Ann. Phys. (N.Y.)} {\bf 149}, 374--456.

\bibitem[\protect\citeauthoryear{Clarke et~al.}{1988}]{ClarkeSci88}
Clarke, J., A.~N. Cleland, M.~H. Devoret, D. Esteve, and J.~M. Martinis: 1988,
  `Quantum mechanics of a macroscopic variable: The phase difference of a
  {J}osephson junction'.
\newblock {\em Science} {\bf 239}, 992--7.

\bibitem[\protect\citeauthoryear{Cleland and
  Geller}{2004}]{Cleland&GellerPRL04}
Cleland, A.~N. and M.~R. Geller: 2004, `Superconducting qubit storage and
  entanglement with nanomechanical resonators'.
\newblock {\em Phys. Rev. Lett.} {\bf 93}, 70501.

\bibitem[\protect\citeauthoryear{Devoret et~al.}{1985}]{DevoretPRL85}
Devoret, M.~H., J.~M. Martinis, and J. Clarke: 1985, `Measurements of
  macroscopic quantum tunneling out of the zero-voltage state of a
  current-biased {J}osephson junction'.
\newblock {\em Phys. Rev. Lett.} {\bf 55}, 1908--11.

\bibitem[\protect\citeauthoryear{Friedman et~al.}{2000}]{FriedmanNat00}
Friedman, J.~R., V. Patel, W. Chen, S.~K. Tolpygo, and J.~E. Lukens: 2000,
  `Quantum superpositions of distinct macroscopic states'.
\newblock {\em Nature (London)} {\bf 406}, 43--6.

\bibitem[\protect\citeauthoryear{Geller and
  Cleland}{2005}]{Geller&ClelandPRA05}
Geller, M.~R. and A.~N. Cleland: 2005, `Superconducting qubits coupled to
  nanoelectromechanical resonators: {A}n architecture for solid-state quantum
  information processing'.
\newblock {\em Phys. Rev. A} {\bf 71}, 32311.

\bibitem[\protect\citeauthoryear{Geller et~al.}{2006}]{GelleretalPre06}
Geller, M.~R., E.~J. Pritchett, A.~T. Sornborger, M. Steffen, and J.~M.
  Martinis: 2006, `Controlled-NOT logic for {J}osephson phase qubits'.
\newblock e-print cond-mat/0000000.

\bibitem[\protect\citeauthoryear{Johnson et~al.}{2003}]{JohnsonPRB03}
Johnson, P.~R., F.~W. Strauch, A.~J. Dragt, R.~C. Ramos, C.~J. Lobb, J.~R.
  Anderson, and F.~C. Wellstood: 2003, `Spectroscopy of capacitively coupled
  {J}osephson-junction qubits'.
\newblock {\em Phys. Rev. B} {\bf 67}, 20509.

\bibitem[\protect\citeauthoryear{Makhlin et~al.}{1999}]{MakhlinNat99}
Makhlin, Y., G. Sch\"on, and A. Shnirman: 1999, `Josephson-junction qubits with
  controlled couplings'.
\newblock {\em Nature (London)} {\bf 398}, 305--7.

\bibitem[\protect\citeauthoryear{Makhlin et~al.}{2001}]{MakhlinRMP01}
Makhlin, Y., G. Sch\"on, and A. Shnirman: 2001, `Quantum-state engineering with
  {J}osephson-junction devices'.
\newblock {\em Rev. Mod. Phys.} {\bf 73}, 357--400.

\bibitem[\protect\citeauthoryear{Marquardt and Bruder}{2001}]{MarquardtPRB01}
Marquardt, F. and C. Bruder: 2001, `Superposition of two mesoscopically
  distinct quantum states: {C}oupling a {C}ooper-pair box to a large
  superconducting island'.
\newblock {\em Phys. Rev. B} {\bf 63}, 54514.

\bibitem[\protect\citeauthoryear{Martinis et~al.}{1985}]{MartinisPRL85}
Martinis, J.~M., M.~H. Devoret, and J. Clarke: 1985, `Energy-level quantization
  in the zero-voltage state of a current-biased {J}osephson junction'.
\newblock {\em Phys. Rev. Lett.} {\bf 55}, 1543--6.

\bibitem[\protect\citeauthoryear{Martinis et~al.}{1987}]{MartinisPRB87}
Martinis, J.~M., M.~H. Devoret, and J. Clarke: 1987, `Experimental tests for
  the quantum behavior of a macroscopic degree of freedom: The phase difference
  across a {J}osephson junction'.
\newblock {\em Phys. Rev. B} {\bf 35}, 4682--98.

\bibitem[\protect\citeauthoryear{Martinis et~al.}{2002}]{MartinisPRL02}
Martinis, J.~M., S. Nam, J. Aumentado, and C. Urbina: 2002, `Rabi oscillations
  in a large {J}osephson-junction qubit'.
\newblock {\em Phys. Rev. Lett.} {\bf 89}, 117901.

\bibitem[\protect\citeauthoryear{Massen van~den
  Brink}{2005}]{MassenvandenBrinkPRM05}
Massen van~den Brink, A.: 2005, `Hamiltonian for coupled flux qubits'.
\newblock {\em Phys. Rev. B} {\bf 71}, 64503.

\bibitem[\protect\citeauthoryear{McDermott et~al.}{2005}]{McDermottSci05}
McDermott, R., R.~W. Simmonds, M. Steffen, K.~B. Cooper, K. Cicak, K.~D.
  Osborn, D.~P. Oh, S.~Pappas, and J.~M. Martinis: 2005, `Simultaneous state
  measurement of coupled {J}osephson phase qubits'.
\newblock {\em Science} {\bf 307}, 1299--302.

\bibitem[\protect\citeauthoryear{Mooij et~al.}{1999}]{MooijSci99}
Mooij, J.~E., T.~P. Orlando, L.~S. Levitov, L. Tian, C.~H. van~der Wal, and S.
  Lloyd: 1999, `Joesphson persistent-current qubit'.
\newblock {\em Science} {\bf 285}, 1036--9.

\bibitem[\protect\citeauthoryear{Nakamura et~al.}{1997}]{NakamuraPRL97}
Nakamura, Y., C.~D. Chen, and J.~S. Tsai: 1997, `Spectroscopy of energy-level
  splitting between two macroscopic quantum states of charge coherently
  superposed by {J}osephson coupling'.
\newblock {\em Phys. Rev. Lett.} {\bf 79}, 2328--31.

\bibitem[\protect\citeauthoryear{Nakamura et~al.}{1999}]{NakamuraNat99}
Nakamura, Y., Y.~A. Pashkin, and J.~S. Tsai: 1999, `Coherent control of
  macroscopic quantum states in a single-{C}ooper-pair box'.
\newblock {\em Nature (London)} {\bf 398}, 786--8.

\bibitem[\protect\citeauthoryear{Orlando et~al.}{1999}]{OrlandoPRB99}
Orlando, T.~P., J.~E. Mooij, L. Tian, C.~H. van~der Wal, L.~S. Levitov, S.
  Lloyd, and J.~J. Mazo: 1999, `Superconducting persistent-current qubit'.
\newblock {\em Phys. Rev. B} {\bf 60}, 15398--413.

\bibitem[\protect\citeauthoryear{Pashkin et~al.}{2003}]{PashkinNat03}
Pashkin, Y.~A., T. Yamamoto, O. Astafiev, Y. Nakamura, D.~V. Averin, and J.~S.
  Tsai: 2003, `Quantum oscillations in two coupled charge qubits'.
\newblock {\em Nature (London)} {\bf 421}, 823--6.

\bibitem[\protect\citeauthoryear{Plastina and Falci}{2003}]{PlastinaPRB03}
Plastina, F. and G. Falci: 2003, `Communicating {J}osephson qubits'.
\newblock {\em Phys. Rev. B} {\bf 67}, 224514.

\bibitem[\protect\citeauthoryear{Pritchett and
  Geller}{2005}]{Pritchett&GellerPRA05}
Pritchett, E.~J. and M.~R. Geller: 2005, `Quantum memory for superconducting
  qubits'.
\newblock {\em Phys. Rev. A} {\bf 72}, 10301.

\bibitem[\protect\citeauthoryear{Rigetti et~al.}{2005}]{RigettiPRL05}
Rigetti, C., A. Blais, and M.~H. Devoret: 2005, `Protocol for universal gates
  in optimally biased superconducting qubits'.
\newblock {\em Phys. Rev. Lett.} {\bf 94}, 240502.

\bibitem[\protect\citeauthoryear{Rouse et~al.}{1995}]{RousePRL95}
Rouse, R., S. Han, and J.~E. Lukens: 1995, `Observation of resonant tunneling
  between macroscopically discinct quantum levels'.
\newblock {\em Phys. Rev. Lett.} {\bf 75}, 1614--7.

\bibitem[\protect\citeauthoryear{Shnirman et~al.}{1997}]{ShnirmanPRL97}
Shnirman, A., G. Sch\"on, and Z. Hermon: 1997, `Quantum manipulations of small
  {J}osephson junctions'.
\newblock {\em Phys. Rev. Lett.} {\bf 79}, 2371--4.

\bibitem[\protect\citeauthoryear{Sornborger et~al.}{2004}]{SornborgeretalPRA04}
Sornborger, A.~T., A.~N. Cleland, and M.~R. Geller: 2004, `Superconducting
  phase qubit coupled to a nanomechanical resonator: {B}eyond the rotating-wave
  approximation'.
\newblock {\em Phys. Rev. A} {\bf 70}, 52315.

\bibitem[\protect\citeauthoryear{Strauch et~al.}{2003}]{StrauchPRL03}
Strauch, F.~W., P.~R. Johnson, A.~J. Dragt, C.~J. Lobb, J.~R. Anderson, and
  F.~C. Wellstood: 2003, `Quantum logic gates for coupled superconducting phase
  qubits'.
\newblock {\em Phys. Rev. Lett.} {\bf 91}, 167005.

\bibitem[\protect\citeauthoryear{van~der Wal et~al.}{2000}]{vanderWalSci00}
van~der Wal, C.~H., A.~C.~J. ter Haar, F.~K. Wilhelm, R.~N. Schouten, C.~J.
  P.~M. Harmans, T.~P. Orlando, S. Lloyd, and J.~E. Mooij: 2000, `Quantum
  superpositions of macroscopic persistent current'.
\newblock {\em Science} {\bf 290}, 773--7.

\bibitem[\protect\citeauthoryear{Vion et~al.}{2002}]{VionSci02}
Vion, V., A. Aassime, A. Cottet, P. Joyez, H. Pothier, C. Urbina, D. Esteve,
  and M.~H. Devoret: 2002, `Manipulating the quantum state of an electrical
  circuit'.
\newblock {\em Science} {\bf 296}, 886--9.

\bibitem[\protect\citeauthoryear{Wallraff et~al.}{2004}]{WallraffNat04}
Wallraff, A., D.~I. Schuster, A. Blais, L. Frunzio, R.-S. Huang, J. Majer, S.
  Kumar, S.~M. Girvin, and R.~J. Schoelkopf: 2004, `Strong coupling of a single
  photon to a superconducting qubit using circuit quantum electrodynamics'.
\newblock {\em Nature (London)} {\bf 431}, 162--7.

\bibitem[\protect\citeauthoryear{Yamamoto et~al.}{2003}]{YamamotoNat03}
Yamamoto, T., Y.~A. Pashkin, O. Astafiev, Y. Nakamura, and J.~S. Tsai: 2003,
  `Demonstration of conditional gate operations using superconducting charge
  qubits'.
\newblock {\em Nature (London)} {\bf 425}, 941--4.

\bibitem[\protect\citeauthoryear{You and Nori}{2005}]{Youreview}
You, J.~Q. and F. Nori: 2005, `Superconducting circuits and quantum
  information'.
\newblock Physics Today, November 2005.
\newblock p. 42.

\bibitem[\protect\citeauthoryear{You et~al.}{2002}]{YouPRL02}
You, J.~Q., J.~S. Tsai, and F. Nori: 2002, `Scalable quantum computing with
  {J}osephson charge qubits'.
\newblock {\em Phys. Rev. Lett.} {\bf 89}, 197902.

\bibitem[\protect\citeauthoryear{Yu et~al.}{2002}]{YuSci02}
Yu, Y., S. Han, X. Chu, S.-I. Chu, and Z. Wang: 2002, `Coherent temporal
  oscillations of macroscopic quantum states in a {J}osephson junction'.
\newblock {\em Science} {\bf 296}, 889--92.

\bibitem[\protect\citeauthoryear{Yukon}{2002}]{Yukon02}
Yukon, S.~P.: 2002, `A multi-{J}osephson junction qubit'.
\newblock {\em Physica C} {\bf 368}, 320--3.

\bibitem[\protect\citeauthoryear{Zhou et~al.}{2004}]{ZhouPRA04}
Zhou, X., M. Wulf, Z. Zhou, G. Guo, and M.~J. Feldman: 2004, `Dispersive
  manipulation of paired superconducting qubits'.
\newblock {\em Phys. Rev. A} {\bf 69}, 30301.

\bibitem[\protect\citeauthoryear{Zhu et~al.}{2003}]{ZhuPRA03}
Zhu, S.-L., Z.~D. Wang, and K. Yang: 2003, `Quantum-information processing
  using {J}osephson junctions coupled through cavities'.
\newblock {\em Phys. Rev. A} {\bf 68}, 34303.

\end{thebibliography}

\end{document}